\documentclass[12pt]{article}

\def\ben{\begin{equation}}
\def\een{\end{equation}}

  \let\n=\nu

\let\C=\Chi

\def\nn{\nonumber} \def\bd{\begin{document}} \def\ed{\end{document}}
\def\ds{\documentstyle} \let\fr=\frac \let\bl=\bigl \let\br=\bigr
\let\Br=\Bigr \let\Bl=\Bigl
\let\bm=\bibitem
\let\na=\nabla
\let\pa=\partial \let\ov=\overline
\newcommand{\be}{\begin{equation}}
\newcommand{\ee}{\end{equation}}
\def\ba{\begin{array}}
\def\ea{\end{array}}
\def\ft#1#2{{\textstyle{{\scriptstyle #1}\over {\scriptstyle #2}}}}
\def\fft#1#2{{#1 \over #2}}
\def\del{\partial}
\def\vp{\varphi}
\def\sst#1{{\scriptscriptstyle #1}}
\def\oneone{\rlap 1\mkern4mu{\rm l}}
\def\td{\tilde}
\def\wtd{\widetilde}
\def\ie{\rm i.e.\ }
\def\dalemb#1#2{{\vbox{\hrule height .#2pt
        \hbox{\vrule width.#2pt height#1pt \kern#1pt
                \vrule width.#2pt}
        \hrule height.#2pt}}}
\def\square{\mathord{\dalemb{6.8}{7}\hbox{\hskip1pt}}}
\newcommand{\ho}[1]{$\, ^{#1}$}
\newcommand{\hoch}[1]{$\, ^{#1}$}
\newcommand{\bea}{\begin{eqnarray}}
\newcommand{\eea}{\end{eqnarray}}
\newcommand{\ra}{\rightarrow}
\newcommand{\lra}{\longrightarrow}
\newcommand{\Lra}{\Leftrightarrow}
\newcommand{\ap}{\alpha^\prime}
\newcommand{\bp}{\tilde \beta^\prime}
\newcommand{\tr}{{\rm tr} }
\newcommand{\Tr}{{\rm Tr} }
\def\0{{\sst{(0)}}}
\def\1{{\sst{(1)}}}
\def\2{{\sst{(2)}}}
\def\3{{\sst{(3)}}}
\def\4{{\sst{(4)}}}
\def\5{{\sst{(5)}}}
\def\6{{\sst{(6)}}}
\def\7{{\sst{(7)}}}
\def\8{{\sst{(8)}}}
\def\n{{\sst{(n)}}}
\def\cA{{{\cal A}}}
\def\cB{{{\cal B}}}
\def\cF{{{\cal F}}}
\def\tV{\widetilde V}
\def\tW{\widetilde W}
\def\tH{\widetilde H}
\def\tE{\widetilde E}
\def\tF{\widetilde F}
\def\tA{\widetilde A}
\def\im{{{\rm i}}}
\def\tY{{{\wtd Y}}}
\def\ep{{\epsilon}}
\def\vep{{\varepsilon}}
\def\R{\rlap{\rm I}\mkern3mu{\rm R}}
\def\bD{{{\bar D}}}

\def\R{\rlap{\rm I}\mkern3mu{\rm R}}
\def\bD{{{\bar D}}}
\def\R{{{\Bbb R}}}
\def\C{{{\Bbb C}}}
\def\H{{{\Bbb H}}}
\def\CP{{{\Bbb C}{\Bbb P}}}
\def\RP{{{\Bbb R}{\Bbb P}}}
\def\Z{{{\Bbb Z}}}
\def\bA{{{\Bbb A}}}
\def\bB{{{\Bbb B}}}
\def\bC{{{\Bbb C}}}
\def\bD{{{\Bbb D}}}
\def\bE{{{\Bbb E}}}
\def\bZ{{{\Bbb Z}}}
\def\Re{{{\frak{Re}}}}
\def\Im{{{\frak{Im}}}}
\def\cosec{{\,\hbox{cosec}\,}}
\def\Gm{{\Gamma_{\!\! -}}}
\def\Gp{{\Gamma_{\!\! +}}}
\def\stan{{standard }}
\def\nonstan{{supernumerary }}

\newcommand{\auth}{H. L\"u\hoch{\dagger} and
J.F. V\'azquez-Poritz\hoch{\ddagger}}

\begin{document}
\begin{flushright}

UK-04-03\ \ \ \ \
UCTP-101-04\\
{\bf hep-th/0401150}\\
January\  2004
\end{flushright}


\begin{center}

{\large {\bf Smooth Cosmologies from M-theory\hoch{\ast}}}

\vspace{20pt}
\auth

\vspace{20pt} {\hoch{\dagger}\it George P. and Cynthia W. Mitchell
Institute for Fundamental Physics,\\ Texas A\& M University,
College Station, TX 77843-4242}

\vspace{10pt} {\hoch{\ddagger}\it Department of Physics and
Astronomy\\ University of Kentucky, Lexington, KY 40506}

\vspace{10pt} {\hoch{\ddagger}\it Department of Physics\\
University of Cincinnati, Cincinnati OH 45221-0011}

\vspace{30pt}

\underline{ABSTRACT}
\end{center}

We review two ways in which smooth cosmological evolution between
two de Sitter phases can be obtained from M/string-theory.
Firstly, we perform a hyperbolic reduction of massive type IIA$^*$
theory to $D=6$ ${\mathcal N}=(1,1)$ $SU(2)\times U(1)$ gauged de
Sitter supergravity, which supports smooth cosmological evolution
between dS$_4\times S^2$ and a dS$_6$-type geometry. Secondly, we
obtain four-dimensional de Sitter gravity with $SU(2)$ Yang-Mills
gauge fields from a hyperbolic reduction of standard
eleven-dimensional supergravity. The four-dimensional theory
supports smooth cosmological evolution between dS$_2\times S^2$
and a dS$_4$-type geometry. Although time-dependent, these
solutions arise from a first-order system {\it via} a
superpotential construction. For appropriate choices of charges,
these solutions describe an expanding universe whose expansion
rate is significantly larger in the past than in the future, as
required for an inflationary model.

{\vfill\leftline{}\vfill \vskip 10pt \footnoterule

{\footnotesize \hoch{\ast} To appear in the Proceedings of the 3rd
International Symposium on Quantum Theory and \phantom{of the}
Symmetries (QTS3) Sept. 10-14, 2003 --- \copyright \ World
Scientific.} \vskip -12pt}  \pagebreak \setcounter{page}{1}

\newpage

\section{$D=6$ ${\mathcal N}=(1,1)$ de Sitter supergravity from massive
type IIA$^*$}

The advantage of embedding de Sitter solutions in $*$ theories
\cite{hull1,hull2} is that they can be viewed as
``supersymmetric,'' even though the anti-commutators of the
super-charges are no longer positive definite \footnote{The first
de Sitter solution within the context of an extended supergravity
theory was found in \cite{gates}.}. The time-like T-dualization of
type IIB theory can yield massive type IIA$^*$ supergravity, whose
bosonic Lagrangian is given by \cite{lvpdstods}
\bea {\mathcal L}_{10}\!\!\! &=&\!\!\!\hat R\, {\hat *\oneone} -
\ft12 {\hat *d\hat \phi}\wedge d\hat \phi + \ft12
e^{\fft32\hat\phi}\, {\hat *\hat F_\2}\wedge \hat F_\2 - \ft12
e^{-\hat \phi}\, {\hat *\hat F_\3}\wedge \hat F_\3\nn\\ &&+ \ft12
e^{\fft12\hat\phi}\, {\hat *\hat F_\4}\wedge \hat F_\4 -\ft12
d\hat A_\3\wedge d\hat A_\3 \wedge \hat A_\2 - \ft16 m\, d\hat
A_\3 \wedge (\hat A_\2)^3\nn\\ && -\ft1{40} m^2 (\hat A_\2)^5
+\ft12 m^2 e^{\fft52\hat \phi}\, {\hat *\oneone},\label{romans1}
\eea
where $\hat F_\2 = d\hat A_\1 + m\, \hat A_\2$, $\hat F_\3 = d\hat
A_\2$ and $\hat F_\4 = d\hat A_\3 + \hat A_\1\wedge d\hat A_\2 +
\ft12 m\, \hat A_\2\wedge \hat A_\2$. The hyperbolic
reduction\footnote{No-go theorems prohibiting de Sitter
compactifications \cite{gibbons,nogo} are surmounted by reducing
over a non-compact space \cite{warner}.} of massive type IIA$^*$
theory can be obtained as an analytical continuation of the $S^4$
reduction \cite{massive} of massive type IIA supergravity
\cite{romans}:
\bea d\hat s_{10}^2&=& c^{\fft1{12}}\, X^{\fft18}\Big[
\Delta^{\fft38}\, ds_6^2 + 2g^{-2}\, \Delta^{\fft38}\, X^2\,
d\xi^2 +\ft12g^{-2}\, \Delta^{-\fft58}\, X^{-1}\, s^2\xi
\sum_{i=1}^3 (h^i)^2\Big]\,,\nn\\
\hat F_\4 &=& -\ft{\sqrt2}{6}\, g^{-3}\, c^{1/3}\, s^3\,
\Delta^{-2}\, U\, d\xi\wedge\ep_\3 -\sqrt2 g^{-3}\, c^{4/3}\,
s^4\, \Delta^{-2}\,
X^{-3}\, dX\wedge \ep_\3 \nn\\
&&-\sqrt2 g^{-1}\, c^{1/3}\, s\, X^4\, {*F_\3}\wedge d\xi
-\ft1{\sqrt2} c^{4/3}\, X^{-2}\, {*F_\2} \nn\\
&& +\ft1{\sqrt2} g^{-2}\, c^{1/3}\, s\, F_\2^i \, h^i\wedge
d\xi\nn\\ && -\ft1{4\sqrt2} g^{-2}\, c^{4/3}\, s^2\, \Delta^{-1}\,
X^{-3}\, F_\2^i \wedge
h^j\wedge h^k\, \ep_{ijk}\,,\label{fans}\\
\hat F_\3 &=& c^{2/3}\, F_\3 + g^{-1}\, c^{-1/3}\, s\, F_\2\wedge
d\xi
\,,\nn\\
\hat F_\2 &=& \ft1{\sqrt2}\, c^{2/3}\, F_\2\,,\qquad e^{\hat\phi}
= c^{-5/6}\, \Delta^{1/4}\, X^{-5/4}\,,\nn \eea
where $X=e^{-\fft1{2\sqrt2}\phi}$, $\Delta \equiv -X\,s^2\xi
+X^{-3}\,c^2 \xi$ and $U \equiv X^{-6}\, c^2 + 3 X^2\, s^2 - 4
X^{-2}\, s^2 - 6 X^{-2}$. We have defined $h^i\equiv \sigma^i-g\,
A_\1^i$, $\ep_\3\equiv h^1\wedge h^2\wedge h^3$, $s=\sinh\xi$,
$c=\cosh\xi$ and $m= \ft{\sqrt2}{3}\, g$. $\sigma_i$ are $SU(2)$
left-invariant 1-forms which satisfy $d\sigma^i = -\ft12
\ep_{ijk}\, \sigma^j\wedge \sigma^k$. $*$ is the six-dimensional
Hodge dual. The resulting six-dimensional theory\footnote{De
Sitter supergravities as hyperbolic reductions of type IIB$^*$ and
M$^*$-theories have recently been obtained in \cite{liu}.} is
given by
\bea {\mathcal L}_6 &=& R\, {*\oneone} -\ft12 {*d\phi}\wedge d\phi
+ g^2\Big(\ft29 e^{\fft{3}{\sqrt2}\phi} -\ft83
e^{\fft1{\sqrt2}\phi} -
2 e^{-\fft1{\sqrt2}\phi}\Big)\,  {*\oneone}\nn\\
&&-\ft12 e^{-\sqrt2\phi}\, {*F_\3\wedge F_\3} +\ft12
e^{\fft1{\sqrt2}\phi}\, \Big( {*F_\2}\wedge F_\2 + {*F_\2^i}\wedge
F_\2^i \Big) \label{d6lag}\\
&& + A_\2\wedge(\ft12 dA_\1\wedge dA_\1 +\ft13 g\, A_\2\wedge
dA_\1 \nn\\ && +\ft2{27} g^2\, A_\2\wedge A_\2 +\ft12 F_\2^i\wedge
F_\2^i)\,,\nn \eea
where $F_\3=dA_\2$, $F_\2= dA_\1 + \ft23g\, A_\2$ and $F_\2^i =
dA_\1^i + \ft12 g\, \ep_{ijk} A_\1^j\wedge A_\1^k$.

Since the theory allows us to truncate to $U(1)^2$, we can include
in our theory a vector-tensor multiplet with the Lagrangian given
by\footnote{Though in general not the case, the truncation is
consistent for the present purpose of constructing cosmological
solutions.}
\be \hat e^{-1}{\mathcal L}_6 = \hat R - \ft12 (\del \phi_1)^2 -
\ft12 (\del \phi_2)^2 - \hat V +\ft14 \sum_{i=1}^2 X_i^{-2}\,
(\hat F_\2^i)^2\,, \ee
where $X_i=e^{\ft12\vec a_i\cdot \vec \phi}$ with $\vec
a_1=(\sqrt2\,, \fft1{\sqrt2})$ and $\vec a_2=(-\sqrt2\,,
\fft1{\sqrt2})$. The scalar potential is given by
\be \hat V=-\ft49 g^2\, (X_0^2 - 9 X_1\, X_2 - 6 X_0\, X_1 - 6
X_0\, X_2)\,, \ee
where $X_0=(X_1\,X_2)^{-3/2}$. We consider the ansatz
\bea
ds_6^2 &=& -d\tau^2 + a^2\, dx_j^2 + b^2\, d\Omega_2^2\,,\nn\\
F_\2^i &=& \lambda_i\, \Omega_\2\,,\label{cosans} \eea
where the functions $a$ and $b$ depend only on the co-moving time
coordinate $\tau$ and $d\Omega_2^2$ is the metric for the unit
2-sphere $S^2$. A cosmological solution can be obtained from the
following first-order equations \cite{lvpdstods}
\bea \dot{\vec \phi} &=& \sqrt2\,\Big(-\fft{1}{2\sqrt2\,g}\,
(q_1\, \vec a_1\, X_1^{-1} + q_2\, \vec a_2\, X_2^{-1})\,
b^{-2} + \fft{dW}{d\vec \phi}\Big)\,,\nn\\
\fft{\dot b}{b} &=& -\fft{1}{4\sqrt2}\,
\Big(\fft3{\sqrt2\,g}\,(q_1\, X_1^{-1} + q_2\, X_2^{-1})\, b^{-2}
+W\Big)\,,\nn\\
\fft{\dot a}{a} &=& \fft{1}{4\sqrt2}\, \Big(
\fft{1}{\sqrt2\,g}\,(q_1\, X_1^{-1} + q_2\, X_2^{-1})\, b^{-2} -
W\Big)\,, \label{6} \eea
provided that the two $U(1)$ charges satisfy
$g\,(\lambda_1+\lambda_2)=1$. The superpotential is given by
$W=\fft{g}{\sqrt2}\,(\ft43 X_0 + 2 X_1 + 2 X_2)$. In particular,
if $\lambda_1=g^{-1}$ and $\lambda_2=0$, then we can consistently
set $\phi_1=2\,\phi_2$ and (\ref{6}) can be solved explicitly.
After the coordinate transformation
$d\tau=\ft32\,e^{\ft{3}{\sqrt{8}}\phi_2} (g\,t)^{-1}\,dt$, the
metric of the solution can be expressed as
\be ds_6^2 = \wtd H^{-\ft14}\, H^{\ft14}\,\Big[-\ft94\, \wtd H\,
H^{-1}\, \fft{dt^2}{(g\, t)^2} +(g\, t)^2\, (dx_i^2 + \wtd H\,
g^{-2}\, d\Omega_2^2) \Big]\,. \label{gendmetricsol} \ee
where
\be e^{\sqrt{8}\, \phi_2}=\fft{\wtd H}{H}=\fft{1 + \ft34\,
\fft{1}{(g\, t)^2}}{1 + \ft94 \, \fft{1}{(g\,
t)^2}}\,.\label{genh} \ee
The solution can be viewed as an intersection of a spatial domain
wall wrapped on $\Omega_2$, characterized by the function $H$, and
an S2-brane characterized by the function $\wtd H$. This is a
smooth cosmological solution in which the co-moving time runs from
an infinite past, which is dS$_4\times S^2$, to an infinite
future, which is a dS$_6$-type geometry with the boundary
$R^3\times S^2$.

For general values of $q_i$, the first-order equations cannot be
solved analytically. However, it is straightforward to find the
fixed-point solution of dS$_4\times S^2$ where $b$ and $\vec \phi$
are constants. Using a numerical method, we have verified that
there are cosmological solutions which run from dS$_4\times S^2$
in the infinite past and are of dS$_6$-type in the future. The
ratio of the Hubble constant in the infinite past and future can
be straightforwardly calculated. In particular, for $-\lambda_1
\gg \lambda_2$, the ratio is given by
\be \fft{H_{{\rm past}}}{H_{{\rm future}}}~
\fft{\sqrt{2}}{16}\Big( \fft{3}{2}\,
\sqrt{-3\fft{\lambda_1}{\lambda_2}}\Big) ^{3/4}, \ee
which can be arbitrarily large. It is surprising that we can get a
somewhat realistic cosmological model from a first-order system,
which implies supersymmetry from the point of view of * theories
\cite{lvpdstods}.

\section{$D=4$ Yang-Mills de Sitter gravity from M-theory}

Four-dimensional de Sitter spacetime arises in standard
eleven-dimensional supergravity as a warped product with a
hyperbolic 7-space \cite{gh}. The latter can be expressed as a
foliation of two 3-spheres, on which $SU(2)$ Yang-Mills fields can
reside. This enables us to use the following reduction ansatz for
$D=11$ supergravity \cite{ds}:
\bea ds_{11}^2 &=& \Delta^{\ft23}\, ds_4^2 + g^{-2}\,
\Delta^{\ft23}\, d\theta^2+ \ft14 g^{-2}\, \Delta^{-\ft13}\, \Big[
c^2\, \sum_i(h^i)^2 + s^2\,\sum_i(\td h^i)^2\Big]
\,,\label{metred}\\
F_\4&=&2g\, \epsilon_\4 -\ft14 g^{-2} \Big(s\,c\, \, d\theta\wedge
h^i\wedge {*F_\2^i} -s\,c\, \, d\theta \wedge \td h^i\wedge
{*F_\2^i}\nn\\
&&- \ft14c^2\, \epsilon_{ijk}\,h^i\wedge h^j\wedge {*F_\2^k} +
\ft14s^2\, \epsilon_{ijk}\, \td h^i\wedge\td h^j\wedge
{*F_\2^k}\Big)\,, \label{f4red} \eea
where $c=\cosh \theta$, $s=\sinh \theta$, $\Delta=\cosh(2\theta)$,
and * denotes the four-dimensional Hodge dual. $\sigma_i$ and $\td
\sigma_i$ are $SU(2)$ left-invariant 1-forms and the vielbein
$h^i$ and $SU(2)$ Yang-Mills field strengths $F_\2^i$ have been
defined in the previous section. It is straightforward to verify
that the Bianchi identity $dF_\4=0$ and the equation of motion
$d{\hat *F_\4}=\ft12 F_4\wedge F_\4$ are satisfied provided that
the $SU(2)$ Yang-Mills fields $A_\2^i$ satisfy the
lower-dimensional equations of motion $D{*F_\2^i}=0$, where the
covariant derivative $D$ is defined by $DV^i=dV^i + g\,
\epsilon_{ijk}\, A^j\wedge V^k$, for any vector $V^i$.

The evaluation of the $D=11$ Einstein equations of motion is
substantially more complicated, and we have not performed the
calculation in full detail.  However, we have verified that the
equations of motion work for the $U(1)$ subsector of the $SU(2)$
gauge fields. Combining the result, the lower-dimensional
equations of motion can be obtained from the Lagrangian
\be e^{-1} {\mathcal L}_4 = R - \ft14 (F_\2^i)^2 - 8
g^2\,.\label{d4lag} \ee
Thus, we have obtained four-dimensional Einstein $SU(2)$
Yang-Mills de Sitter gauged gravity from $D=11$ by consistent
Kaluza-Klein reduction on a hyperbolic 7-space \cite{ds}.

A regular cosmological solution of (\ref{d4lag}) is given by
\cite{adsbh}
\bea ds_4^2 &=& H^2\, (-f^{-1}\, dt^2 + f\, dx^2 +
t^2\, d\Omega_2^2)\nn\\
F_\2^3 &=& \fft{2\ell}{(t\,H)^2}\, dt\wedge dx\,,\qquad
{*F_\2^3}=2\ell\, \, \Omega_\2\,,\nn\\
H&=&1 +\fft{\ell}{t}\,,\qquad f=\ft43g^2 t^2\, H^4 -1\,. \eea
For $g^2\,\ell^2 = \ft{3}{64}$, this is smooth cosmological
evolution between dS$_2\times S^2$ and a dS$_4$-type geometry with
the boundary of $S^2\times S^1$. It is straightforward to lift the
solution back to $D=11$ and obtain a regular cosmological solution
in M-theory.  The corresponding metric is given by
\bea ds_{11}^2 &=& \Delta^{\ft23}\, H^2\, \Big( -f^{-1}\, dt^2 +
f\, dx^2 + t^2\, d\Omega_2^2 \Big) + g^{-2}\,\Delta^{\ft23}\,
d\theta^2\nn\\ && +\ft14g^{-2}\, \Delta^{-\ft13}\, \Big[
c^2\,\Big(d\omega_2^2 + (\sigma_3 -\fft{2g}{H}\,dx)^2\Big)
\\ && + s^2\,\Big(d\wtd\omega_2^2 + (\td\sigma_3
-\fft{2g}{H}\,dx)^2\Big)\Big]\,.\nn \eea

\section{Discussion}

The crucial issue regarding both of the above cosmological
solutions is their stability. Although the * theories are
necessary from the point of view of time-like T-duality, they
suffer from an instability due to the ghost-like nature of the
supergravity fields. It is of interest, therefore, to further
study whether the stability is protected by time-like T-duality or
``supersymmetry.'' Our second example might be more appealing
since it is embedded within standard eleven-dimensional
supergravity. The optimistic view is that the time scale of the
instability is large enough to, nevertheless, validate the
cosmological solutions.

Metastable de Sitter vacua have recently been constructed in type
IIB theory \cite{kklt}. All moduli are frozen due to fluxes as
well as corrections to the superpotential from Euclidean D-brane
instantons or gaugino condensation. However, these techniques can
only yield inflationary models under restrictive conditions
\cite{kklmmt}. If such moduli-freezing techniques can be
incorporated into our smooth inflationary models, then this might
greatly enhance the cosmological landscape of M-theory
\cite{susskind}. In particular, we would have cosmologies which
smoothly evolve (without topological transitions) between de
Sitter-type spacetimes of different dimensionalities.

\section{Acknowledgements}

J.F.V.P. thanks the organizers of the 3rd International Symposium
on Quantum Theory and Symmetries (QTS3) held at the University of
Cincinnati. Research supported in part by DOE grant
DE-FG03-95ER40917 (H.L.) and DOE grant DE-FG01-00ER45832
(J.F.V.P.).

\end{document}